# Modeling fragmentation of the self-gravitating molecular layer by smoothed particle hydrodynamics

M. Nejad-Asghar[1] and D. Molteni[2]

[1] Department of Physics, Damghan University of Basic Sciences, Damghan, Iran, email: nasghar@dubs.ac.ir
[2] Dipartimento di Fisica e Tecnologie Relative, Universita di Palermo, Viale delle Scienze, 90128 Palermo, Italy

**Abstract.** We revisit the modeling of ion-neutral (or ambipolar) diffusion with two fluid smoothed particle hydrodynamics, as discussed by Hosking & Whitworth. Some parts of the technique are optimized to testify the pioneer works on behavior of the ambipolar diffusion in an isothermal self-gravitating layer. The frictional heating by ambipolar diffusion is examined, and its effect on fragmentation of the layer is studied. The results are compared to the thermal phases of instability as obtained by Nejad-Asghar.

**1 Introduction**

Ambipolar diffusion was first proposed in an astrophysical context by Mestel & Spitzer (1956) as a mechanism for removing magnetic flux and hence magnetic pressure from collapsing protostellar cores. Likewise, ion-neutral friction has long been considered to be a source of heating mechanism in molecular clouds (e.g., Scalo 1977). Padoan et al. (2000) show that ambipolar drift heating is a strong function of position in a turbulent cloud, and its average value can be significantly larger than the average cosmic ray heating rate. Ambipolar diffusion has recently been invoked as a source of thermal instability in molecular clouds for formation of small-scale condensations (Nejad-Asghar 2007).

Many authors have developed computer codes that attempt to model ambipolar diffusion. Black & Scot (1982) used a two-dimensional, deformable-grid algorithm to follow the collapse of isothermal, non-rotating magnetized cloud. The three-dimensional work of MacLow et al. (1995) treats the two-fluid model in a version of the *zeus* magnetohydrodynamic code. An algorithm capable of using the smoothed particle hydrodynamics (SPH) to implement the ambipolar diffusion in a fully three-dimensional, self-gravitating system was developed by Hosking & Whitworth (2004, hereafter HW). They described the SPH implementation of two-fluid technique that was tested by modeling the evolution of a dense core, which is initially thermally supercritical but magnetically subcritical.

In this paper, we optimize the two-fluid SPH implementation of the ambipolar diffusion in one dimensional layer. Formulation of the ambipolar diffusion in an isothermal self-gravitating layer is obtained in section 2. We re-formulate the optimized two-fluid SPH technique in section 3. Furthermore, we deal with some prospects of the fragmentation of the layer via thermal instability.

**2 Isothermal molecular layer**

Considering a plane-parallel self-gravitating layer of lightly ionized isothermal molecular gas, all variables are functions of distance $z$ to the central plane and time $t$ only. We consider the molecular cloud as global neutral which consists of a mixture of atomic and molecular hydrogen, helium, and traces of other rare molecules. In reality, the gas in this case is very weakly ionized. The magnetic field is frozen only to the ions, so that diffusion of the field relative to the neutral gas must continuously insert a drag force (per unit volume)

$$f_d = \gamma_{AD}\rho_i\rho v_d, \qquad (1)$$

where $\gamma_{AD} \approx 3.5\times 10^{10}$ m$^3$/kg.s represents the collision drag coefficient and $\rho_i$ is the ion density. The exact fraction of the total fluid that is ionized depends upon many factors (e.g. the neutral density, the cosmic ray ionization rate, how efficiently ionized metals are depleted on to dust grains). Many analytical models have been developed to determine the variation of the ion density with neutral density by considering ionization balance in typical molecular cloud environments. Here, we use the expression employed by Fiedler & Mouschovias (1992), which states that for $10^8 < n < 10$ m$^{-3}$,

$$\rho_i = \varepsilon(\rho^{1/2} + \varepsilon'\rho^{-2}), \qquad (2)$$

where in standard ionized equilibrium state, $\varepsilon \approx 7.5 \times 10^{-15}$ kg$^{1/2}$.m$^{-3/2}$ and $\varepsilon' \approx 4 \times 10^{-44}$ kg$^{5/2}$.m$^{-15/2}$ are valid.

It is a good approximation to treat the charged particles as a single fluid, therefore, the drift velocity is given by

$$v_d = -\frac{1}{\gamma_{AD}\varepsilon\rho^{3/2}(1+\varepsilon\rho^{-5/2})}\frac{\partial}{\partial z}(\frac{B^2}{2\mu_0}), \qquad (3)$$

which is obtained by assumption that the pressure and gravitational force on the charged fluid component are negligible compared to the Lorentz force because of the low ionization fraction. In this way, the magnetic fields are directly evolved by charged fluid component, as follows:

$$\frac{dB}{dt} = -B\frac{\partial v}{\partial z} + \frac{\partial}{\partial z}(Bv_z). \qquad (4)$$

The ion density is negligible in comparison to the neutral density, thus, the mass conservation is given by

$$\frac{d\rho}{dt} = -\rho\frac{\partial v}{\partial z}. \qquad (5)$$

The momentum equation then becomes

$$\frac{dv}{dt} = g - \frac{1}{\rho}\frac{\partial}{\partial z}(p + \frac{B^2}{2\mu_0}) \qquad (6)$$

where $p = a^2\rho$ is the pressure ($a$ is the isothermal sound speed) and the gravitational acceleration $g$ obeys the poisson's

$$\frac{\partial g}{\partial z} = -4\pi g\rho. \qquad (7)$$

Following the many previous treatments, a further simplification is possible if we introduce the surface density between mid-plane and $z > 0$ as

$$\sigma \equiv \int_0^z \rho(z',t)dz'. \qquad (8)$$

By transformation from $(z,t)$ to $(\sigma,t)$, the drift velocity is given by

$$v_d = -\frac{1}{\gamma_{AD}\varepsilon\rho^{1/2}(1+\varepsilon\rho^{-5/2})}\frac{\partial}{\partial \sigma}(\frac{B^2}{2\mu_0}), \qquad (9)$$

and the equation (4) becomes

$$\frac{\partial}{\partial t}(\frac{B}{\rho}) = \frac{1}{\gamma_{AD}\varepsilon\mu_0}\frac{\partial}{\partial \sigma}[\frac{B^2}{\rho^{1/2}(1+\varepsilon'\rho^{-5/2})}\frac{\partial B}{\partial \sigma}]. \qquad (10)$$

With the above, field equation (7) can be integrated to give

$$g = -4\pi G\sigma, \qquad (11)$$

while the equation of continuity (5) and the equation of motion (6) take the form

$$\frac{\partial z}{\partial \sigma} = \frac{1}{\rho} \qquad (12)$$

and

$$\frac{\partial^2 z}{\partial t^2} = -4\pi G\sigma - \frac{\partial}{\partial \sigma}(a^2\rho + \frac{B^2}{2\mu_0}), \qquad (13)$$

respectively. The layer is assumed to be in quasi-magnetohydrostatic equilibrium at all times, supported against its own self-gravity by the magnetic gas pressure. The loss of flux from ambipolar diffusion is exactly compensated for by the compression of the layer which is necessary to maintain equilibrium. In this approximation, the left-hand side of equation (13) is zero, and we may integrate the force balance to obtain

$$\frac{B^2}{2\mu_0} + a^2\rho = 2\pi G(\sigma_\infty^2 - \sigma^2) \qquad (14)$$

where integration constant $\sigma_\infty$ is the value of $\sigma$ at $z = \infty$ (where $\rho$ is zero). Following the work of Shu (1983), we introduce the non-dimension quantities

$$\tilde{\sigma} \equiv \frac{\sigma}{\sigma_\infty}, \quad \tilde{\rho} \equiv \frac{a^2}{2\pi G \sigma_\infty^2}\rho, \quad \tilde{z} \equiv \frac{2\pi G \sigma_\infty}{a^2}z, \quad \tilde{B} \equiv \frac{B}{2\sigma_\infty \sqrt{\pi \mu_0 G}},$$

$$\tilde{t} \equiv (\frac{2\sqrt{2\pi G}}{\gamma_{AD}\varepsilon})(\frac{2\pi G \sigma_\infty}{a})t, \quad \tilde{\varepsilon} \equiv \frac{a^5}{(2\pi G)^{5/2}\sigma_\infty^5}\varepsilon', \quad \tilde{v}_d \equiv \frac{\gamma_{AD}\varepsilon}{a\sqrt{2\pi G}}v_d, \tag{15}$$

so that we rewrite the basic equations (10), (12) and (14) as follows:

$$\frac{\partial}{\partial \tilde{t}}(\frac{\tilde{B}}{\tilde{\rho}}) = \frac{\partial}{\partial \tilde{\sigma}}(\frac{\tilde{B}^2}{\tilde{\rho}^{1/2}+\tilde{\varepsilon}\tilde{\rho}^{-2}}\frac{\partial \tilde{B}}{\partial \tilde{\sigma}}), \tag{16}$$

$$\frac{\partial \tilde{z}}{\partial \tilde{\sigma}} = \frac{1}{\tilde{\rho}}, \tag{17}$$

$$\tilde{B}^2 + \tilde{\rho} = 1 - \tilde{\sigma}^2, \tag{18}$$

and the drift velocity (9) as

$$\tilde{v}_d = -\frac{1}{\tilde{\rho}^{1/2}+\tilde{\varepsilon}\tilde{\rho}^{-2}}\frac{\partial \tilde{B}^2}{\partial \tilde{\sigma}}. \tag{19}$$

Here, a natural family of initial states is generated by assuming that the initial ratio of magnetic to gas pressure is everywhere a constant, $\alpha_0$, i.e., $\tilde{B}^2/\tilde{\rho} = \alpha_0$ at $\tilde{t}=0$. Then one finds from equations (17), (18) and (19) that

$$\rho_{(z,t=0)} = \frac{\rho_0}{\cosh^2(z/z_\infty)}, \tag{20}$$

$$v_{d(z,t=0)} = \frac{2\alpha_0}{\sqrt{1+\alpha_0}}\frac{a\sqrt{2\pi G}}{\gamma_{AD}\varepsilon}\frac{\sinh(z/z_\infty)}{1+\varepsilon'\rho_0^{-5/2}\cosh^5(z/z_\infty)}, \tag{21}$$

where $\rho_0 \equiv 2\pi G \sigma_\infty^2/a^2(1+\alpha_0)$ is the central density of the layer at $t=0$, and $z_\infty \equiv a\sqrt{(1+\alpha_0)/2\pi G\rho_0}$ is a length-scale parameter. Figure 1 shows the initial neural density, ion density, and drift velocity in the cloud and outercloud medium for $\rho_0 = 1.7\times10^{-14}\,\text{kg.m}^{-3}$, $a = 0.3\,\text{km.s}^{-1}$, and $\alpha_0 = 1$.

We can solve the equations (16)-(19) by finite difference techniques under the initial and boundary conditions. Figure 2a presents graphically the solution for $\tilde{v}_d$, which carried out to time $\tilde{t}=20$ with initial condition $\alpha_0 = 1$. As the magnetic field leaks from the neutral gas, the volume density of the neutrals shifts in profile from equation (20) to that case with $\alpha_0 = 0$ (Fig. 2b), and the drift velocity gradually settles.

## 3 Two-fluid SPH construction

In two-fluid SPH technique of HW, the initial SPH particles are represented by two sets of *molecular particles*: magnetized ion SPH particles and non-magnetized neutral SPH particles. In this method, for each SPH particle we must create two separate neighbor lists: one for neighbors of the same species and another for those of different species. Consequently, each particle must have two different smoothing lengths. In the following sections we refer to neutral particles as $\alpha$ and $\beta$, and ion particles as $a$ and $b$, the subscripts 1 and 2 refer to both ions and neutral particles. We adopt the usual smoothing spline-based kernel (Monaghan & Lattanzio 1985) and apply the symmetrized form proposed by Hernquist & Katz (1989). We update the adaptive smoothing length according to the octal-tree based nearest neighbor search algorithm (NNS).

The chosen physical scales for length and time are $[l]=200\,\text{AU}$, and $[t]=10^3\,\text{yr}$, respectively, so that velocity unit is approximately $[v]=1\,\text{km.s}^{-1}$. The gravity constant is set $G=1[m]^{-1}[l]^3[t]^{-2}$ for which the calculated mass unit is $[m]=4.5\times10^{29}\,\text{kg}$. Consequently, the derived physical scale for density and drag coefficient are $[\rho]=1.7\times10^{-11}\,\text{kg.m}^{-3}$ and $\gamma_{AD}=1.8\times10^{10}[l]^3[m]^{-1}[t]^{-1}$, respectively. In this manner, the numerical values of $\varepsilon$ and $\varepsilon'$ are $1.8\times10^{-9}[l]^{-3/2}[m]^{1/2}$ and $3.5\times10^{-17}[l]^{-15/2}[m]^{5/2}$, respectively. The magnetic field is scaled in units such that the constant $\mu_0$ is unity. Specifying $\mu_0 = 1$ therefore scales the magnetic field equal to $[B]=5.1\,\text{nT}$.

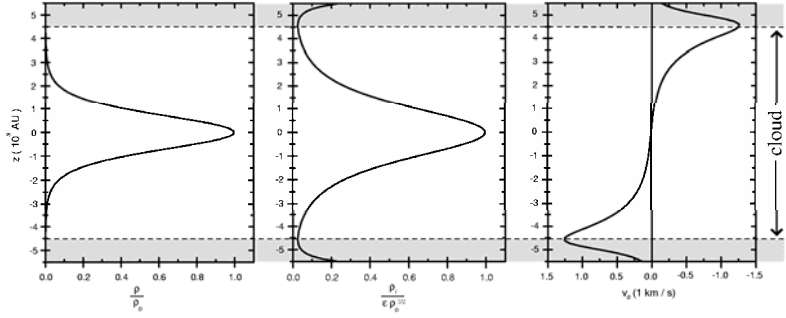

**Fig. 1.** The initial neural density, ion density, and drift velocity in the cloud and outercloud medium (gray region) for $\rho_0 = 1.7 \times 10^{-14}$ kg.m$^{-3}$, $a = 0.3$ km.s$^{-1}$, and $\alpha_0 = 1$.

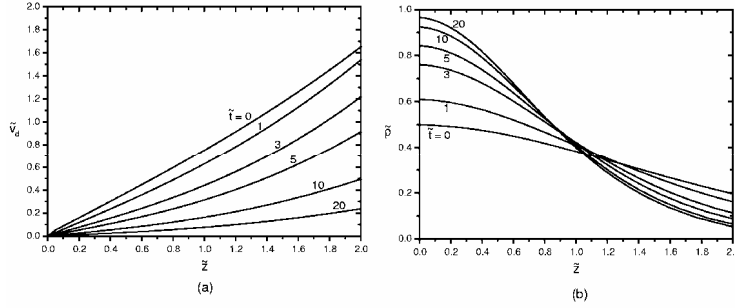

**Fig. 2.** (a) The non-dimensional drift velocity plotted against the non-dimensional vertical coordinate at times $\tilde{t}$ equal to 0, 3, 5, 10, and 20 using the finite difference technique. (b) The same plot as (a) but for non-dimensional neutral gas density.

The initial conditions for this simulation are a density given by equation (20), with a parallel magnetic field directed perpendicular to the $z$-axis so that the initial ratio of magnetic to gas pressure is everywhere a constant ($\alpha_0 = 1$). The central density is assumed to be $10^{-3}[\rho]$ and the magnetic field is assumed to be frozen in the fluid of charged particles. We choose a molecular cloud which the mean molecular mass of neutrals and ions are respectively $2.3 m_H$ and $30 m_H$, where $m_H$ is the mass of hydrogen atom. Both fluids are assumed to be isothermal, and at a sound speed of $0.3[v]$.

The neutral density in place of neutral particles, $\rho_{n,a}$, is estimated via usual summation over neighboring neutral particles, while in place of ions, $\rho_{n,a}$, is given by interpolation technique from the values of the nearest neighbors. The ion density is evaluated via equation (2) for both places of ions and neutral particles. According to this new ion density, we update the mass of ion $a$ as

$$m_a^{new} = m_a^{old} \frac{\rho_a^{new}}{\rho_a^{old}}, \tag{22}$$

in each time step so that the usual summation law for density of ions might being appropriate.

The SPH form of the drift velocity of ion particle $a$ is given by HW as

$$v_{d,a} = \frac{1}{\gamma_{AD}\rho_{n,a}}[-\frac{1}{\mu_0 \rho_{i,a}}\sum_b \frac{m_b}{\rho_{i,b}} B_b B_a \frac{dW_{ab}}{dz_a} - \sum_b m_b \Pi_{ab}\frac{dW_{ab}}{dz_a}] \tag{23}$$

where $\Pi_{ab}$ is the artificial viscosity between ion particles $a$ and $b$. But, it is better to remember the second golden rule of SPH which is to rewrite formulae with the density inside operators (Monaghan 1992). In this case, the drift velocity (3) is interpolated as

$$v_{d,a} = \frac{1}{\gamma_{AD}\rho_{n,a}}[-\frac{1}{2\mu_0 \rho_{i,a}}\sum_b \frac{m_b}{\rho_{i,b}}(B_b^2 - B_a^2)\frac{dW_{ab}}{dz_a} - \rho_{i,a}\sum_b \frac{m_b}{\rho_{i,b}}\Pi_{ab}\frac{dW_{ab}}{dz_a}], \tag{24}$$

where two extra density terms are introduced, one outside and one inside the summation sign. This comes as a result of the approximation to the volume integral needed to perform function interpolation.

There is not any analytical expression that lets us calculate the value of drift velocity of the neutral particles. Here, we use the interpolation technique that starts at the nearest neighbor, then add a sequence of decreasing corrections, as information from other neighbors is incorporated (e.g., Press et al. 1992). These drift velocities at neutral places are used to estimate the drag acceleration, $a_{drag} = \gamma_{AD}\rho_{i,\alpha}v_{d,\alpha}$, instead the method of HW who used the expression of Monaghan & Kocharayan (1995).

The gravitational acceleration in the self-gravitating SPH acceleration equation for neutral particle $\alpha$ may be obtained by equation (11), as follows

$$a_{grav,\alpha} = \pm 4\pi G \sigma_{\alpha} \quad (25)$$

where plus is for $z_{\alpha} < 0$, minus is for $z_{\alpha} > 0$, and $\sigma_{\alpha}$ is the total surface density between mid-plane and neutral particle $\alpha$. The ion momentum equation assuming instantaneous velocity update so that we have

$$v_a = \sum_{\beta} \frac{m_{\beta}}{\rho_{\beta}} v_{\beta} W_{\alpha\beta} + v_{d,a} \quad (26)$$

where the first term on the right-hand side gives the neutral velocity field at the ion particle $a$, calculated using the standard SPH approximation.

In ambipolar diffusion process, the ion particles are physically diffused through the neutral fluid, thus, the ions will be bared in the boundary regions of the cloud (i.e. without any neutral particles in their neighbors). We check the position of ions before making a tree and nearest neighbor search, so that we send out the bared ion particles in the boundary regions of the simulation at next time-step. We assume that the cloud layer is spread from $z = -4.5z_{\infty}$ to $z = +4.5z_{\infty}$ (according to Fig. 1). Both the cloud and boundary regions contain ion and neutral particles. The complete system is represented by $N$ ($=10^3$) discrete but smoothed SPH particles (i.e. Lagrangian sample points) with $N/2$ ions and $N/2$ neutral particles. We set up boundary particles ($4h_1$ up and down in $z$) using the linear extrapolation approach (from the values of the inner particles) to attribute the appropriate drift velocity, drag acceleration, pressure acceleration, and the magnetic induction rate to the boundary particles.

The present SPH code has the main features of the TreeSPH class (Barnes & Hut 1986), so that the nearest neighbors searching are calculated by means of this procedure. We integrate the SPH equations using the simple integration scheme. The selection of time-step, $\Delta t$, is of great importance. There are several time-scales that can be defined locally in the system. For each particle 1, we calculate the smallest of these time-scales using its smallest smoothing length, $h_1$, i.e.

$$\Delta t_1 = C_{cour} \min[\frac{h_1}{|v_1|}, \frac{h_1}{v_{A,1}}, \frac{h_1}{a_1}], \quad (27)$$

where $v_A = B/\sqrt{\mu_0 \rho_i}$ is the Alfven speed of ion fluid and $C_{cour}$ is the Courant number which in this paper is adopted equal to 0.8 (for numerical stability). The evolution were carried out to time $56[t]$ (equivalent to $\tilde{t} = 1$), a chore that took about 10 hours on a 3.6 GHz (Pentium IV) processor. Figure 3 presents the diffusion of the magnetic field profile which accurately matches with Fig. 2 of this paper and/or Figure 1 of Shu (1983).

The energy equation follows from the first law of thermodynamics, that is

$$\frac{1}{\gamma-1}\frac{dp}{dt} - \frac{\gamma}{\gamma-1}\frac{p}{\rho}\frac{d\rho}{dt} + \rho\Omega_{(\rho,T)} = 0, \quad (28)$$

where $\gamma$ is the polytropic index, and $\Omega_{(\rho,T)}$ is the net cooling function. Nejad-Asghar (2007) has considered a constant initial form of the temperature profile in the layer, and has solved numerically the nonlinear isobaric energy equation with approximation that the dynamics of the layer is solely determined by the isothermal equations as outlined by Shu (1983). In this way, a large temperature gradient was created with time at the interfaces between different isobaric unstable thermal phases. In the context of his thermal instability analysis in the molecular layer, he found that the isobaric thermal instability can occur in some regions of it; therefore it may produce the layer fragmentation and formation of the spherical AU-scale condensations. Trusting on the two-fluid SPH code, we can go ahead to study the thermal phases of the aforementioned molecular layer. Running this process by smoothed particle hydrodynamics is a future work of the authors. Furthermore, the macro-velocity fields in the molecular clouds are highly turbulent and supersonic (Larson 1981). It is then of uppermost importance to consider both collision and merger of the formed condensations. Merging is possibly the main onset mechanism to form dense cores, which likely evolves to star formation.

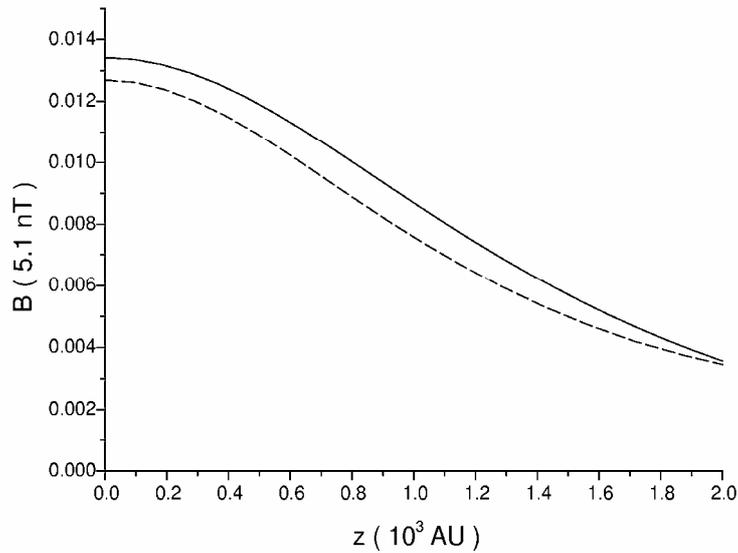

**Fig. 3.** The magnetic field against the vertical distance at times $t=0$ (solid) and $5.65 \times 10^4$ yr (dash), using the two-fluid SPH technique.